\title{Determination of the best optimal estimation parameters for
validation of infrared 
hyperspectral sounding retrievals}

\author{X. Calbet\\
EUMETSAT, Am Kavallerisand 31, 64295 Darmstadt, Germany\\
(Xavier.Calbet@eumetsat.int)}

\date{\today}

\documentclass[a4paper,12pt]{article}
\usepackage{graphicx}

\begin{document}
\maketitle

\begin{abstract}
The availability of hyperspectral infrared remote sensing
instruments, like AIRS and IASI, on board of Earth observing satellites 
opens the possibility of obtaining high vertical resolution
atmospheric profiles. We present an objective and simple technique
to derive the parameters used in the optimal estimation
method that retrieve atmospheric states from the spectra.
The retrievals obtained in this way are optimal in the sense
of providing
the best possible validation statistics obtained
from the difference between retrievals and
a chosen calibration/validation dataset of atmospheric states.
This is demonstrated analytically.
To illustrate this result
several real world examples using IASI retrievals
fine tuned to ECMWF analyses
are shown.
The analytical equations obtained
give further insight into the various contributions 
to the biases and errors of the
retrievals and the consequences of using other types of fine tuning.
Retrievals using IASI
show an error of $0.9$ to $1.9\; {\rm K}$ 
in temperature and
below $6.5\; {\rm K}$ in humidity dew point temperature
in the troposphere on the vertical radiative transfer
model pressure grid (RTIASI-4.1), which
has a vertical spacing between 300 and 400 m.
The more accurately the calibration dataset represents the
true state of the atmosphere, the better the retrievals
will be when compared to the true states.
\end{abstract}

%
%

%

%
%

\section{Introduction}

Temperature and water vapor soundings from satellites has a history dating
back to the early 1970s with the NIMBUS series of operational weather
satellites (e.g. Wick, 1971). The next generation of instruments improved
the horizontal and vertical resolution of the soundings, in particular
the instruments that comprise the TOVS (TIROS Operational Vertical Sounder; 
Smith et al., 1979) and the more recent
ATOVS (Advanced TIROS Operational
Vertical Sounder; Kidwell, 1986). The latter consists of 
the AMSU (Advanced Microwave Sounding Unit)
and HIRS (High Resolution Infrared Sounder)
and provides soundings with an accuracy of about $2.0 \rm{K}$ for the temperature
at 1--km vertical resolution and below $6.0 \rm{K}$ for the dewpoint temperature at
2--km vertical resolution (Li, 2000).
However, in order to make further advancements, the numerical weather prediction and climate monitoring communities
required improvements in both accuracy  for temperature ($< 1 \rm{K}$)
and  for humidity ($< 10\%$) in the troposphere (World Meteorological
Organization, 1998). It became apparent that to achieve these accuracies a
new generation of instruments known now as hyperspectral infrared sounders
were needed. Smith (1991) gives a detailed overview of the evolution
of satellite sounding up to the hyperspectral sounders we have today.
AIRS (Atmospheric InfraRed Sounder; Pagano et al. 2003; Aumann et al. 2003)
and IASI (Infrared Atmospheric Sounding Interferometer; Chalon, Cayla and Diebel 
2001; Blumstein et al. 2004),
on board of Earth observing satellites are the prime
examples of these type of instruments with a spectral coverage within
the $3.62-15.5 {\rm \mu m}$ region and a spectral resolution ($\lambda/\Delta \lambda$)
higher than $1000$ which can achieve the required accuracies in temperature
and humidity retrievals in the troposphere
(Smith 1991).

There are currently three types of methods generally used
to retrieve temperature and humidity profiles
from these type of instruments: 
linear regression methods
usually based on Empirical Orthogonal Functions
(e.g. Smith and Woolf, 1975; Zhou, 2002),
neural networks (e.g. Blackwell, 2005) and
model inversion methods (Twomey, 1977), also known as Bayesian 
atmospheric flux inversion (Michalak, 2005), optimal estimation (Rodgers, 1976),
 physical retrievals (Li, 2000; Susskind, 2003).
The latter type of methods can vary in several details such as
the choice of measurement error covariance matrix, constraints applied, etc.
Whenever one of these methods makes use of synthetic radiances
generated from a radiation transfer model, some inversion parameters
must be fine tuned to match the retrieval method to the real world
measurements.
In this paper, we will deal only with the
model inversion method which is commonly known as
optimal estimation (Rodgers, 2000).
Its particularity is that the constraints are based on 
an a--priori state of the atmosphere and its associated background
covariance matrix.
The same type of analytical study as the one 
presented here can be made on
the EOF linear regression method when it is trained with
synthetic data (Calbet and Schl\"{u}ssel 2006) or on
any of the other methods as long as they can be described analytically.

Methods to retrieve temperature and humidity profiles
from these kind of instruments have been developed
over the years. Modern methods can perform retrievals
over clear, cloudy, land or ocean scenes (Susskind, 2003;
Zhou, 2005). Despite the significat abundance of non--clear
cases, the physical state parameters are better known
for clear over ocean scenes, in particular surface
emissivity and cloud properties. In fact, these cases provide the best
retrieval statistics (e.g. Susskind 2003).
Because of this, the calibration explained in this paper
is preferably best performed on these cases.
This fine tuning can later be extrapolated
to retrieve profiles for any kind of scene. Also, to prove that the fine
tuning derived here is optimal we need to verify it
in practice with the best possible cases available. This is the reason
why in this paper we will deal mainly with clear sky over
ocean scenes.

The first and most critical tuning step is to adjust
the numerical modelling of the atmosphere to the real
world. This can be done by fine tuning the radiative transfer
model to fit the measurements (Strow, 2006).
This procedure is usually not enough to obtain un--biased
retrievals and it is usually necessary to apply bias corrections to the 
radiances or
brightness temperature
(Li, 2000; Susskind, 2006).
The bias corrections are usually obtained
from the measurements by calculating the average of
the difference between 
the real observed spectra and the calculated spectra obtained
from some collocated calibration dataset of atmospheric states 
and a radiative transfer model.
In a later retrieval step, measured (or modeled) spectra must be bias corrected
with the above calculated average.
In this paper we will provide an analytical justification to use
this method and we will also show that it is optimal.

The second set of elements to be tuned in the retrieval process 
are the various parameters
used in the optimal estimation: measurement error covariance matrix, a--priori
state and a--priori error covariance matrix.
For the retrieval method to work precisely we need to match
with a relatively high degree of accuracy
these elements of the retrievals to the real atmospheric system.
The a--priori parameters are usually determined from some
climatology or numerical weather model fields
which are representative of the atmospheric states being
retrieved. Contrary to the bias corrections,
the measurement error covariance matrix
is not usually derived from the measurements, but rather
taken from the instrument noise
and/or adding to it some estimation of the radiative transfer model
error (e.g. Susskind, 2003; Rodgers, 1990; Eyre 1990). The reason for this
is to maintain the measurement error covariance as small
as possible to obtain some ``ideal best'' retrieval,
relying very much on the assumptions made and the validity of the
modelling of the atmosphere.
In this paper we will abandon this hypothetical concept
of ``ideal best'' retrievals and take a more pragmatic
approach. We will aim to reproduce as accurately as possible
the states of the atmosphere as measured by some
alternative instrument, typically either radiosondes or Numerical
Weather Prediction (NWP) analyses. These measurements will be denominated
calibration or validation dataset indistinctively throughout this paper.
By minimizing the standard
deviation between the retrievals and the validation dataset
of atmospheric states we will demonstrate analytically that there
exists an optimal
measurement error covariance matrix for a particular
validation dataset.
This measurement error covariance matrix
is precisely the one obtained
from the difference between 
the real observed spectra and the calculated spectra obtained
from the calibration dataset of atmospheric states and 
the radiative transfer model.
Since this matrix will also contain representativeness and
accuracy errors of the calibration dataset of atmospheric states,
it will depart from an hypothetical ``ideal best'' measurement error.
We will show that this is normally not a problem in the retrievals since
it is safer to overestimate (Eyre, 1990)
the measurement error of the retrievals with respect to some absolute truth,
as it certainly happens with the method presented here,
than to underestimate it, as it
can potentially happen by using the above mentioned methods.

In the field of trace gases retrievals, 
Michalak et al. (2005) are also deriving the measurement error
covariance matrix from the measurements themselves.
The technique consists of maximizing the likelihood of the
measurement covariance given a radiative transfer model,
an a--priori state and a given set of measurements by applying
the Bayes' rule. The maximizing solution is quite complex
analytically and it has to be solved numerically with an iterative
process to obtain the measurement error covariance. 
The spirit of the method presented here is very similar
to the one from Michalak et al. (2005) but minimizing
the statistics of the retrievals with respect to the
validation dataset of atmospheric states, therefore obtaining the optimal
retrievals.
They are optimal in the sense that they give
the minimum bias and standard deviations when compared
to the validation dataset.
Being a simple method, the optimal measurement error covariance 
matrix can be derived analytically and calculated with
real data in a straight forward way. The method offers an 
objective methodology for populating
the measurement error covariance of the optimal estimation.

The fact that the solution is derived analytically gives
an important insight into what elements are affecting
the resulting statistics of the
retrievals when compared with the validation dataset of atmospheric
states. We can even see how the retrieval error behaves when
we use different measurement error covariance matrices.
Other behaviors of the retrieval system could be further studied,
like for example what are the consequences of using another optimal
criteria different than minimizing the statistics of the retrievals,
etc.

In Section \ref{sec:method} the method and the underlying assumptions
are explained. The analytical results and proof of this method being optimal
are explained in Section \ref{sec:optimal} and
shown in Appendix \ref{sec:proof}. The tools used to apply the method are explained
in Section \ref{sec:example}.  Results of this method applied to real
world data are shown in Section \ref{sec:results}.
 Finally in Section \ref{sec:discussion} 
we discuss the method and results.

\section{Best Parameter Determination Method}
\label{sec:method}

\subsection{General Assumptions}

There are some underlying assumptions when applying this method
which are worth mentioning. They are directly related to the
goals we are pursuing with the retrievals.

\begin{enumerate}
 \item
The ultimate goal are the retrievals themselves
and to have
them as accurate as possible. This
also implies that we want to validate
the retrievals with an alternative measurement of the
state of the atmosphere.

\item
We recognize that the current modelling of the atmosphere
is not accurate enough to provide some ``ideal best''
retrievals, but rather that we will need to calibrate
the whole retrieval system with some calibration dataset
of atmospheric states.

\item
We will assume that the scene under observation is measured
only once from space, although more than one instrument or channel
can be used. In this paper we will use IASI measurements.
See Toohey and Strong (2007) for an interesting discussion
on cross calibration of different platforms.

\item
We have only one alternative type of measurement of the atmospheric
state
which will constitute our calibration/validation dataset.
For example,
in this paper we will use NWP analyses fields to calibrate
and optimize the validation of the retrievals.
This is in contrast to using more than one
source of atmospheric information, like
combining NWP and radiosonde data to fine tune and optimize the retrieval
parameters.
Dealing with two different sources of atmospheric measurements
for fine tuning
will not be dealt with in this paper. Even so, we can still perform
the exercise of validating the retrievals with another source
of atmospheric information as is shown in Section \ref{sec:sondes} with
radiosondes.

\end{enumerate}

\subsection{Fine Tuning and Retrieval}

The method used can be divided in two steps, the fine
tuning one and the retrieval proper one.
These in turn can be broken down in the following substeps,

\begin{enumerate}
 \item
Tuning step.

\begin{enumerate}
\item
In the tuning step
the measurements, radiances or brightness temperature, from the hyperspectral
instrument
are obtained.
These will be referred to as
observations (OBS).

\item
The next step is to find the co--located calibration dataset of 
the atmospheric state vector
corresponding to the same scene as the IASI observation.
We then
calculate the spectra corresponding to this atmospheric state
vector using a radiative transfer model.
These will be referred to as calculations (CALC).

\item
We then calculate the difference between observations and calculations
(OBS - CALC) for many different scenes.

\item
The last step is to get the statistics of this collection of OBS - CALC.
In particular the mean (bias) and the covariance
matrix.

\item
Optimal estimation is designed to work with data that
has Gaussian noise. This requirement has to be fulfilled
by the OBS-CALC difference. This working hypothesis should be verified
by checking the OBS-CALC histograms for each channel or measurement.

\end{enumerate}

\item
Retrieval step.

\begin{enumerate}
\item
In the retrieval step the measured radiances or brightness
temperatures are corrected with the bias
calculated in the fine tuning step.

\item
We then use the OBS--CALC covariance obtained in the
fine tuning step directly as the measurement error covariance matrix of the optimal estimation.

\item
The background state vector and its covariance matrix
can be calculated from any source as long as their statistics
are similar to the ones from the real atmosphere.
In this particular example they have been obtained from
climatology.

\end{enumerate}

\end{enumerate}

\section{Best parameter determination for the optimal estimation}
\label{sec:optimal}

\subsection{Bias corrections}

In Appendix \ref{sec:proof} we give the analytical proof that
the fine tuning method described here is the optimal one.
It is derived for the linear method but can also be applied
to the non--linear case if the first guess is close enough
to the final result.

The biases in the retrievals are given by Eq. \ref{eq:biasexpression},
which we replicate here,

\begin{eqnarray}
 \overline{x_R  - x_v} = 
 \left[ K^T S_{\epsilon}^{-1} K + S_a^{-1} \right]^{-1} \cdot \nonumber \\
 \left[
K^T S_{\epsilon}^{-1} ( \overline{y_o - y_c} ) +
S_a^{-1} ( \overline{x_a - x_v}) \right],
\label{eq:biasexpression2}
\end{eqnarray}

where $x_R$ is the retrieved atmospheric state, $x_v$ is the atmospheric
state from the calibration
or validation dataset, $K$ is the Jacobian from the radiative transfer, 
$S_a$ is the background covariance matrix, $S_\epsilon$ is
the measurement error covariance matrix, $y_o$ is the observed (OBS) spectrum,
$y_c$ is the calculated (CALC) spectrum and $x_a$ is the background a--priori
atmospheric state.
We can see from this expression that the bias comes from two sources. The first
source is the OBS - CALC bias in the spectra ($\overline{y_o - y_c}$).
The second one comes from the difference between the background a--priori
 and the validation dataset atmospheric state.
The latter error should be small if the information content of the
radiance spectra is high as is the case for IASI.
In order to eliminate these biases in the retrievals,
the spectral
measurements should be bias corrected according to Eq. \ref{eq:bias}:

\begin{equation}
 \overline{y_c} = \overline{y_o},
\end{equation}

which is effectively an OBS--CALC bias correction.
Also the a--priori state, which is a constant in the
retrievals, should match the calibration dataset states average
(Eq. \ref{eq:xa}),

\begin{equation}
\label{eq:xa2}
 x_a = \overline{x_v}.
\end{equation}

\subsection{Measurement and background error covariances}

In an ideal or simulated world the measurement error covariance matrix,
$S_{\epsilon,i}$, is quite accurately defined as

\begin{equation}
S_{\epsilon,i} = \overline{ (y_i - F_p(x_t) (y_i - F_p(x_t) )^T},
\end{equation}

where $y_i$ is an idealised instrument spectrum, 
$F_p$ represents a perfect  radiative
transfer model and $x_t$ is the true atmospheric state of the atmosphere.
However, in the real world the instrument does not behave ideally, the
radiative transfer model is not perfect and the true atmospheric state
can only be approximated by measurements. This in turn implies that
there is no practical way to derive the ideal $S_{\epsilon,i}$.
A practical alternative solution 
is to estimate it from simultaneous measurements
of the atmospheric state and spectra,

\begin{equation}
S_{\epsilon} = \overline{ (y_o - F(x_v) (y_o - F(x_v) )^T},
\end{equation}

where $y_o$ is the observed spectrum, 
$F$ represents the radiative model used
and $x_v$ is the measurement of the state of the atmosphere.
Obviuosly, the better each of the components that go into this
equation are, the better the approximation of $S_{\epsilon,i}$ will be.
Hopefully the instrument should be well behaved, the radiative transfer
model should reproduce the radiation properly and the measured atmospheric
states should be as representative of the true state as possible.

In what follows we will show that this latter $S_\epsilon$ is actually
the one that minimizes the errors of the
retrievals when compared to the validation dataset. For practical
purposes, this value is a good estimation of the ideal
covariance, $S_{\epsilon,i}$, as will be shown
in later sections with real world examples and by showing below
that it is better to overestimate the measurement errors than to underestimate
them in the retrievals.

\begin{figure}
\noindent\includegraphics[angle=-90,width=\textwidth]{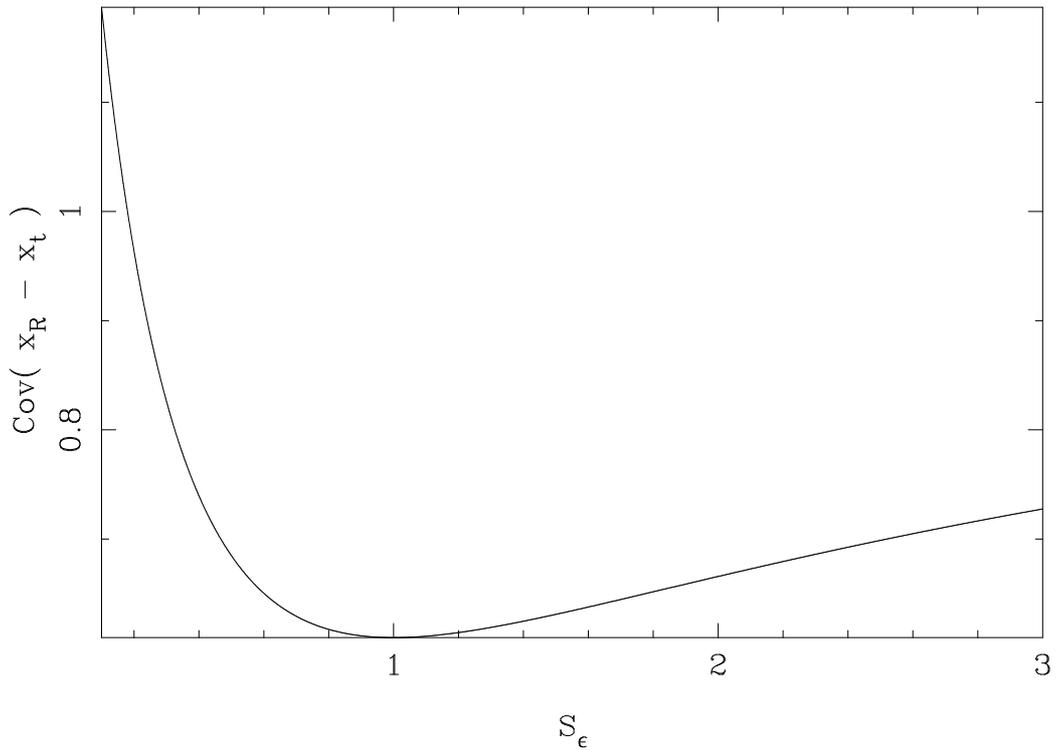}
\caption{\label{fig:safe}
Retrieval error (as the covariance of the difference
between the retrieved parameter and the real one, ${\rm Cov}( x_R - x_v ) = \overline{(x_R-x_v)^2}$)
as a function of the measurement error covariance, $S_\epsilon$,
as described by Eq. \ref{eq:cov1d}.
Other values used in this plot are
$K = 0.8$; $S_a = {\rm Cov}(x_a - x_v) = \overline{(x_a-x_v)^2} = 1.0$ and
${\rm Cov}( y_o - y_c ) = \overline{(y_o - y_c)^2} = 1.0$.
}
\end{figure}

The covariance of the retrieval error is shown in
Eq. \ref{eq:covariance}. To see what effect the different
values of the measurement error has on a particular retrieval
system, we can simplify the expression to just one measurement
and one retrieved variable. This expression will not show
in full detail how the actual retrieval with many variables
behaves, and it is just illustrative,

\begin{equation}
\label{eq:cov1d}
{\rm Cov}( x_R - x_v ) = \frac{K^2 S_\epsilon^{-2} \overline{(y_o - y_c)^2} + S_a^{-2} \overline{( x_a - x_v )^2} }
{(K^2 S_\epsilon^{-1} + S_a^{-1})^2}.
\end{equation}
We can now plot the retrieval error as a function of $S_\epsilon$ for a given
set of parameters. This is shown in Fig. \ref{fig:safe}.
We can see clearly that the retrieval error increases much more rapidly when
we underestimate the measurement error than when we overestimate it. Assuming,
just for this argument, that $x_v$ is actually the absolute true state
of the atmosphere and not a calibration dataset as in the rest of this paper, we
can see that
for practical purposes it is ``safer'' to overestimate the measurement error than to underestimate it.

The measurement error covariance
matrix, $S_{\epsilon}$, that minimizes the errors of the retrievals 
with respect to the validation dataset
is found analytically to be (Eq. \ref{eq:se}),
\begin{equation}
 S_{\epsilon} = \overline{(y_o - y_c) (y_o - y_c)^T},
\end{equation}
which is exactly the OBS--CALC covariance matrix.
To minimize the resulting retrieval errors, 
the a--priori covariance matrix, $S_a$, should satisfy (Eq. \ref{eq:sa}),
\begin{equation}
 S_a = \overline{(x_a - x_v)(x_a - x_v)^T}.
\end{equation}
This expression, together with Eq. \ref{eq:xa2}, basically states that the
a--priori covariance matrix should match the covariance matrix of the validation
atmospheric states.

One important aspect of this analytical proof is that it can be
easily modified to be used with other retrieval methods or
even for validation parameters other than the average or
standard deviation.

\section{Practical Example}
\label{sec:example}

\subsection{IASI Infrared Hyperspectral Measurements}

The real world measurements come from the IASI instrument.
IASI is a hyperspectral resolution infrared sounder on board of the polar orbiting series of Metop satellites that forms the EUMETSAT Polar System (EPS). Metop-A, the first of three satellites of the series was launched successfully on 19 October 2006, from the Baikonur Cosmodrome in Kazakhstan. IASI is a Michelson interferometer measuring between 3.62 and 15.5 microns with a spectral resolution of $0.5\; {\rm cm}^{-1}$ after apodisation. The spatial resolution is of 12 km at nadir.

\subsection{Scene Selection}
\label{sec:scene}

The scenes observed by IASI were selected
for the fine tuning and retrieval step
as clear sky over ocean at nighttime with latitudes equatorward
of $50^\circ$. The reason
for this is to keep to a minimum unknown effects
which might show up, like for example,
unknown surface emissivity over land, cloud properties, etc.
The selection criteria to declare a certain scene
cloudy or clear in the fine tuning is very critical. If a small
percentage of the scenes are cloud contaminated, this
will lead to a bigger than desired bias in the final validation
of the retrievals.
In this paper the clear scenes selection method
is the one followed by Lutz (2002, 2003) and is shown tabulated
in Table \ref{tab:scene}.
A total of 5308 scenes have been selected
around midnight and noon on the days of 10, 11, 17, 18,
19, 27, 28 and 29 of April 2007.
They were selected plus or minus one hour from midnight or noon to have
an NWP analysis field close enough in time to the
IASI observations.
This sample was split in two, a first one of 5042 scenes
to calculate the fine tuning coefficients and the rest 266
to be validated against NWP analyses fields.
Although the scenes used to validate the retrievals are
also clear sky over ocean ones,
the same fine tuning could be used on any other type of scene.
An example of this is shown in Section \ref{sec:sondes} where the retrievals
are compared with co--located radiosondes.

\subsection{Radiative Transfer Model}

The radiative transfer model used is RTIASI 4.1 (Matricardi and
Saunders 1999).
This fast model provides both direct
radiances or brightness temperatures for each IASI channel and their
corresponding Jacobians.
RTIASI also has a built-in model of surface emissivity
which we have used in practice.

\subsection{Optimal Estimation Retrievals}

\begin{figure}
\noindent\includegraphics[angle=-90,width=\textwidth]{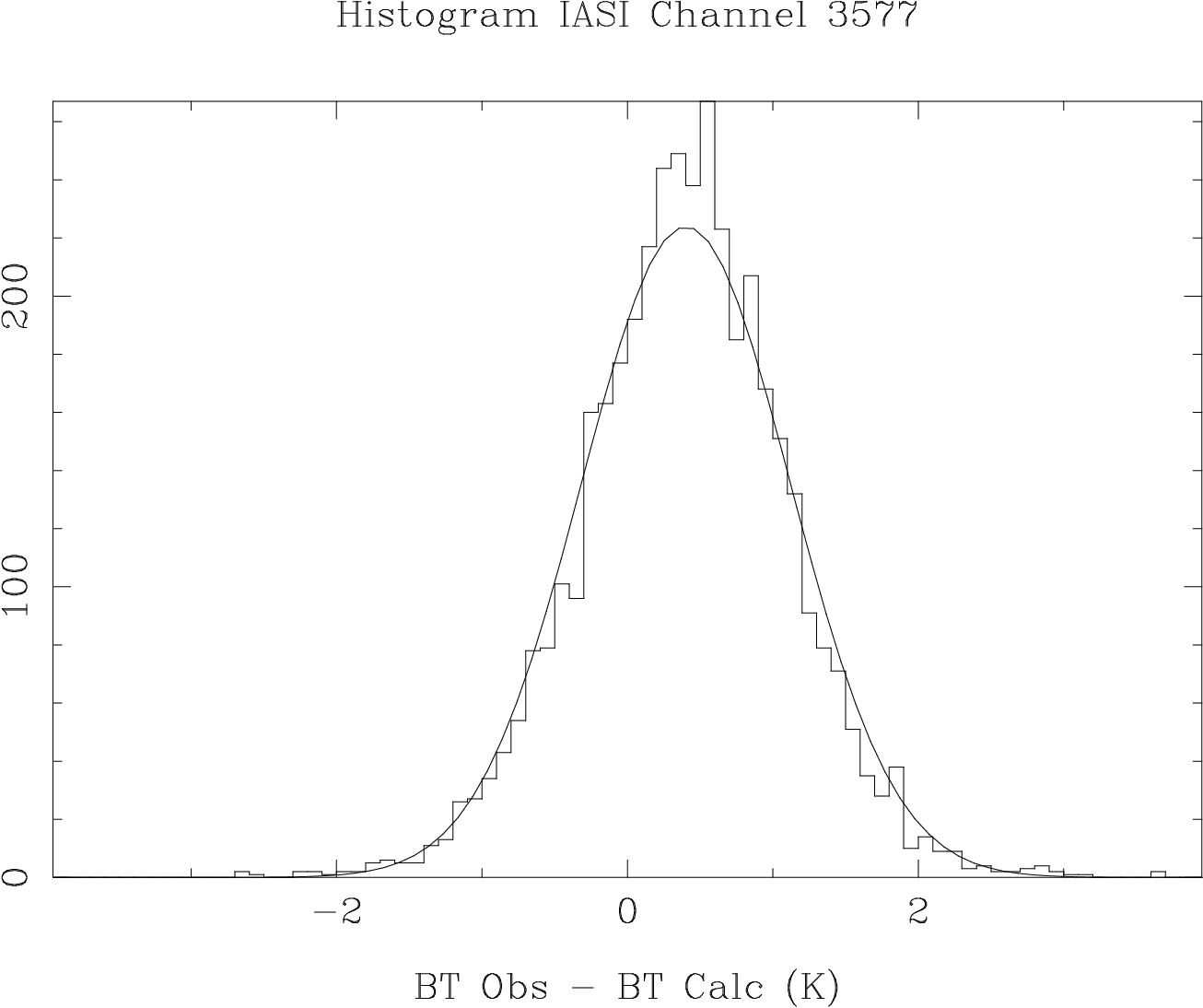}
\caption{\label{fig:gaussian}
Histogram of brightness temperature difference between
observed and calculated spectra (OBS--CALC) for IASI channel 3577 
($1539 \; {\rm cm^{-1}}$).
Stepwise line is the measured histogram and smooth line is the fitted Gaussian.}
\end{figure}

\begin{figure}
\noindent\includegraphics[angle=-90,width=\textwidth]{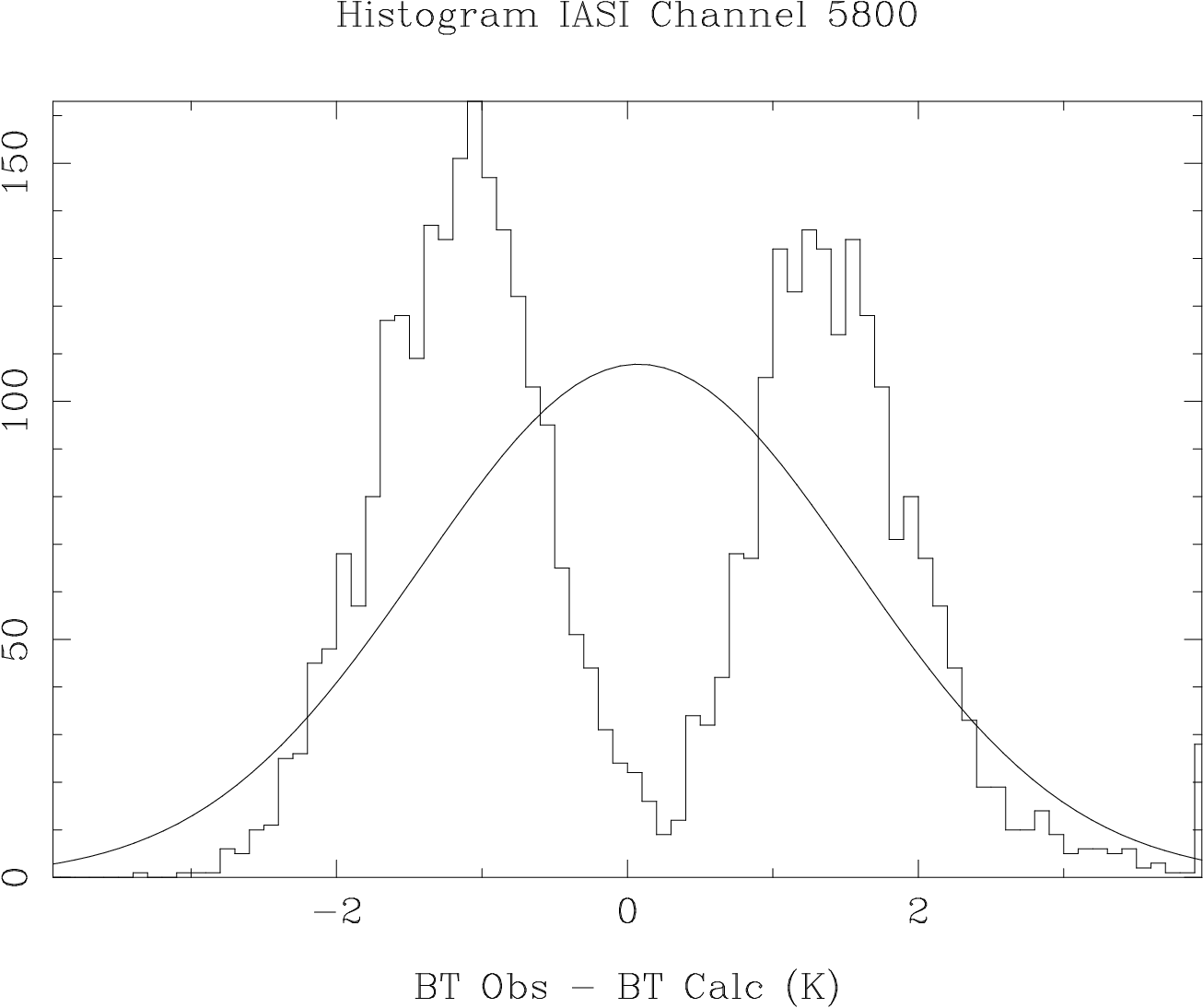}
\caption{\label{fig:nongaussian}
Histogram of brightness temperature difference between
observed and calculated spectra (OBS--CALC) for IASI channel 5800
($2094.75 \; {\rm cm^{-1}}$).
Stepwise line is the measured histogram and smooth line is the fitted Gaussian.}
\end{figure}

The retrieval method used is the non--linear optimal estimation
one as explained in Rodgers (2000).
The technique is applied in brightness temperature space.
One of the pre-requisites to apply this method
is that the errors are Gaussian.
Although the instrument
error is Gaussian only in radiance space, the
global
error covariance matrix from OBS--CALC in brightness temperature space
is in fact also Gaussian.
Indeed, since the OBS--CALC error covariance matrix includes,
besides instrument error, also radiative model errors
and NWP errors, the overall effect
is a Gaussian error in brightness temperature space.
This can be verified in Fig. \ref{fig:gaussian} for channel
3577 ($1539 \; {\rm cm^{-1}}$) 
where the histogram of the OBS--CALC brightness temperature
has been plotted.
As a counter--example and for illustrative purposes, we also 
show the histogram for channel
5800 ($2094.75 \; {\rm cm^{-1}}$) in Fig. \ref{fig:nongaussian}, which clearly
deviates from a Gaussian function. This anomaly comes from
an incorrect CO input profile to the radiative transfer model.
To solve this problem we have to either discard this channel,
which is the solution adopted in this paper, or try to introduce
a more realistic CO profile.

Optimal estimation retrievals are performed on the temperature and
water vapor profiles and skin temperature.
First guess estimates come from a previous
EOF linear regression retrieval (Calbet and Schl\"{u}ssel, 2006)
of ozone, temperature and water vapor profiles
and skin temperature.

The channels used in the retrievals are the ones
with wavenumbers smaller than $1900\,{\rm cm}^{-1}$, except the ones on the ozone
band. The reasons for avoiding the shortwavelength region is
that it is difficult to model daytime radiation effects, the instrument
noise is high and there are absorption lines of some trace gases from
which the atmospheric profiles are difficult to know (e.g. CO).
The ozone band is not used because the ozone profile
is not retrieved in the optimal estimation.

Brightness temperatures are bias corrected with
the OBS--CALC obtained from the fine tuning step.
They are then used by
the optimal estimation retrieval.

The measurement error covariance matrix of the optimal estimation
is the square of the standard deviation of OBS--CALC.
Although the optimal error covariance matrix
is actually the full OBS--CALC covariance matrix (Eq. \ref{eq:se}), it has been
verified by our own experience
that only the diagonal (i.e. square of the standard deviation) 
is actually needed in these particular retrieval exercises. This
slightly simplifies the procedure.

The atmospheric state vectors used to calculate
the a--priori parameters (a--priori state vector
and a--priori covariance matrix) are a modified
subset of the Chevallier profiles (Chevallier 2002).
These profiles constitute a representative sample
of the atmosphere obtained from the 40--year
re--analysis project of the European Centre
for Medium--Range Weather Forecasts (ECMWF).

The non--linear optimal estimation method is solved
iteratively using a minimization Levenberg--Marquardt
algorithm. The iterations are finished when the cost
function does not decrease significantly anymore.
Note that a consequence of this is finalizing
the retrievals with brightness temperature residuals
well below the $1-\sigma$ level of the
measurement error covariance matrix, $S_\epsilon$.
See Section \ref{sec:discussion} for a more in depth discussion.

\subsection{Calibration/Validation Dataset of Atmospheric States}

The reference states of the atmosphere will be the
NWP analyses from ECMWF. They are co--located
by choosing the atmospheric profile of the
nearest grid point of the ECMWF analysis
to the IASI field of view. They also are at most only one hour
apart from the IASI measurement.

\subsection{Instrument Noise}

Instrument noise does not
have a Gaussian behavior in brightness temperature space,
which is what is required by optimal estimation.
Nevertheless, for the purpose of comparing with the optimal
error covariance matrix (OBS--CALC), retrievals
were done using instrument noise as the sole
contribution to the measurement error covariance matrix.
In these cases, brightness temperature instrument
noise was calculated based on its measured brightness temperature
for each IASI field of view.

\section{Practical Example Results}
\label{sec:results}

\subsection{Best Parameter Determination Method Results}

\begin{figure}
\noindent\includegraphics[angle=-90,width=\textwidth]{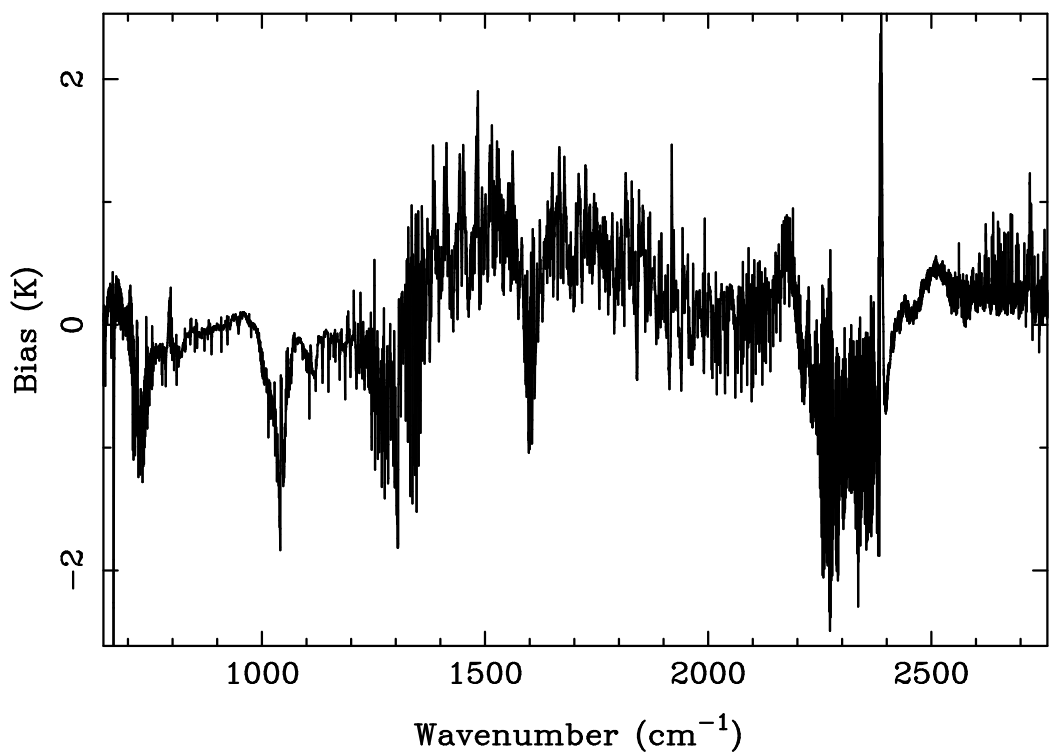}
\caption{\label{fig:bias}
OBS--CALC bias: observed minus calculated (ECMWF analyses + RTIASI 4.1)
brightness temperature averages for all IASI channels.}
\end{figure}

\begin{figure}
\noindent\includegraphics[angle=-90,width=\textwidth]{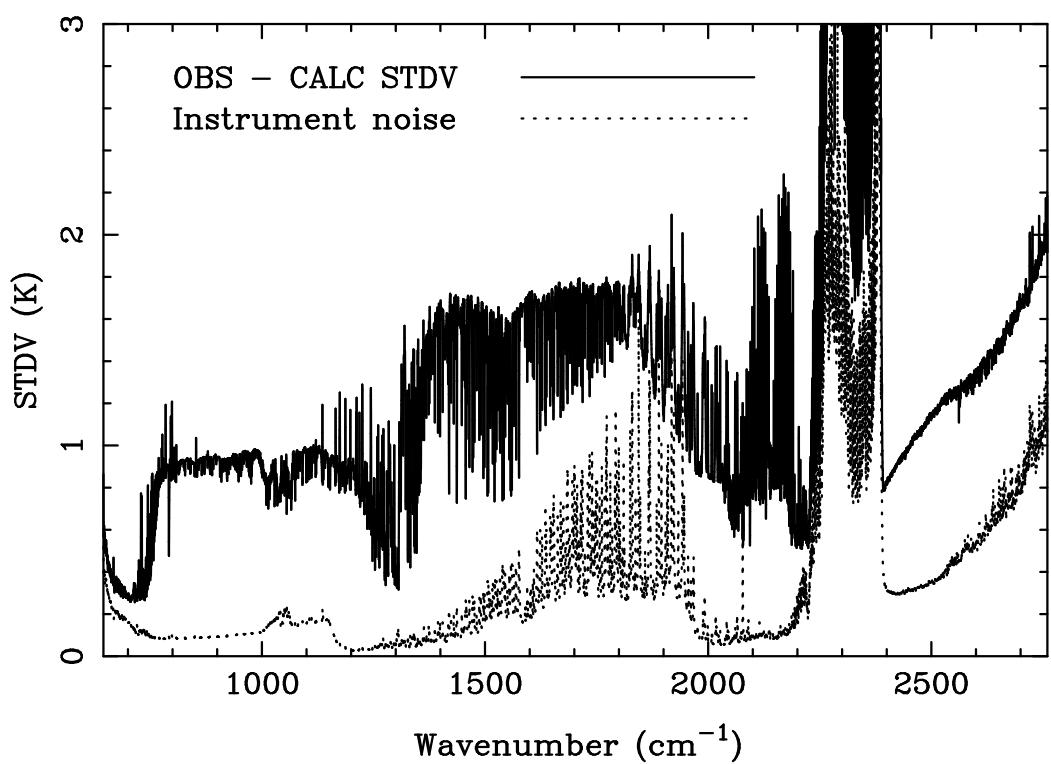}
\caption{\label{fig:stdv}
OBS-CALC standard deviation: observed minus calculated (ECMWF analyses
+ RTIASI 4.1) brightness temperature standard deviations for all IASI channels.
Also shown is the instrument noise for one randomly chosen atmospheric
state.}
\end{figure}

The OBS--CALC statistics for the 5042 profiles are shown in Fig. \ref{fig:bias} and \ref{fig:stdv}.
Fig. \ref{fig:bias} shows the bias for each of the IASI wavelengths.
The standard deviation of OBS--CALC as a function of IASI
wavelegnth is shown in Fig. \ref{fig:stdv}. For comparison purposes,
the instrument noise in brightness temperature space for one randomly chosen
IASI spectrum is also shown in Fig. \ref{fig:stdv}.
We can see how the total error in some regions is much higher
than the instrument noise. In those particular channels
where this is the case, the error contribution from
the radiative transfer modelling or the representativeness
of ECMWF analyses is much higher than the instrument noise.
Note that since we are stopping the iterations of the optimal estimation
algorithm when the cost function does not descend significantly,
the final brightness temperature residuals are well below
the values of the
measurement error covariance matrix, $S_\epsilon$.
See Section \ref{sec:discussion} for a more in depth discussion.

\subsection{Effects of Different Measurement Error Covariance Matrices}

In order to illustrate that the OBS--CALC covariance matrix
is effectively the optimum one to use for the retrievals,
three different error covariance matrices have been applied:
the optimum one (OBS--CALC), a constant standard
deviation of $2 {\rm K}$ and using only the instrument noise.
The retrieval technique is the optimal
estimation explained in Section \ref{sec:method} for all three
experiments.
Retrievals where made on the 266 measured IASI fields of view
(which are independent of the 5042 scenes used for fine tuning). In this
case non-polar, clear air, nighttime  over the ocean
retrievals were performed (see Section \ref{sec:scene}).

\begin{figure}
\noindent\includegraphics[angle=-90,width=\textwidth]{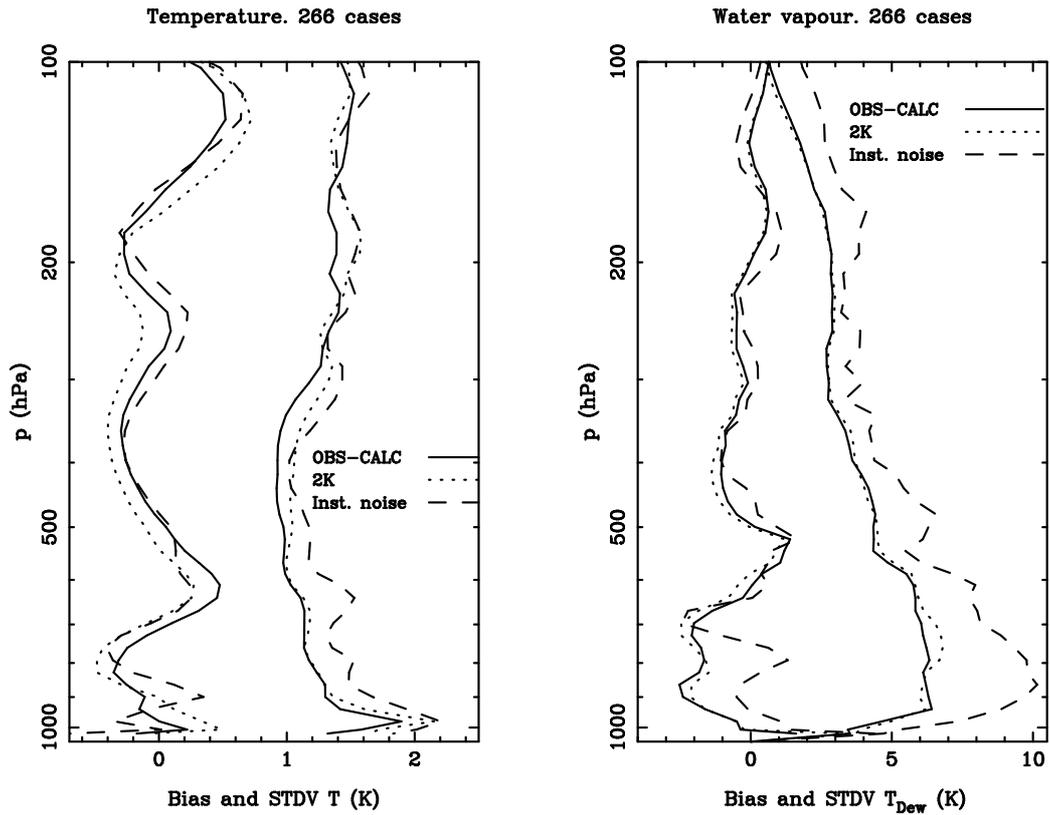}
\caption{\label{fig:stat_all}
Bias (curves on the left of each graph) and standard deviation 
(curves on the right of each graph) of the retrieval statistics using the diagonal of the instrument noise, a constant of $2\;{\rm K}$ and OBS--CALC standard deviation as the error
covariance matrix in the retrievals. The error covariance matrix that provides
the optimal retrievals when comparing with ECMWF analyses is the OBS--CALC one
as the analytical proof of Appendix \ref{sec:proof} shows.}
\end{figure}

A comparison of all three methods (OBS--CALC, $2 {\rm K}$ and instrument noise
 as measurement error covariance matrices) 
can be seen in Fig. \ref{fig:stat_all}.
As was expected, optimum covariance matrix (OBS--CALC) offers the
best retrievals within the statistical noise of the comparison.

\subsection{Statistics and Examples of Optimum Retrievals}

\begin{figure}
\noindent\includegraphics[angle=-90,width=\textwidth]{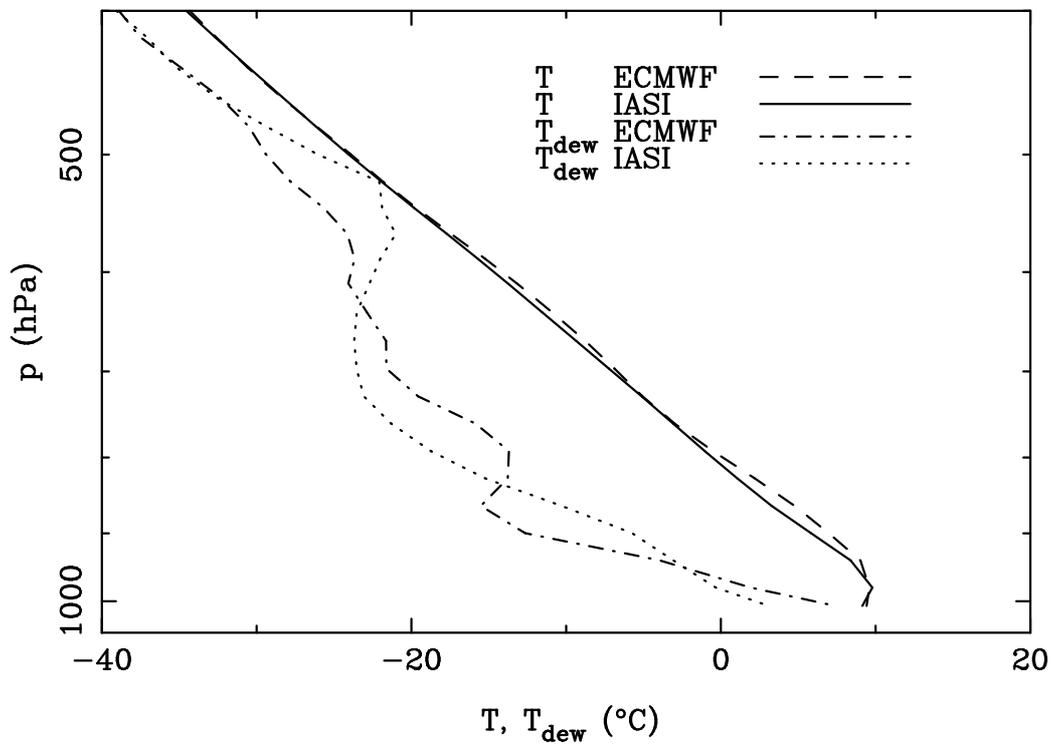}
\caption{\label{fig:oe2221}
Typical IASI retrieval is shown together with the
co--located ECMWF atmospheric profile. There is a low level inversion
that is clearly retrieved.}
\end{figure}

\begin{figure}
\noindent\includegraphics[angle=-90,width=\textwidth]{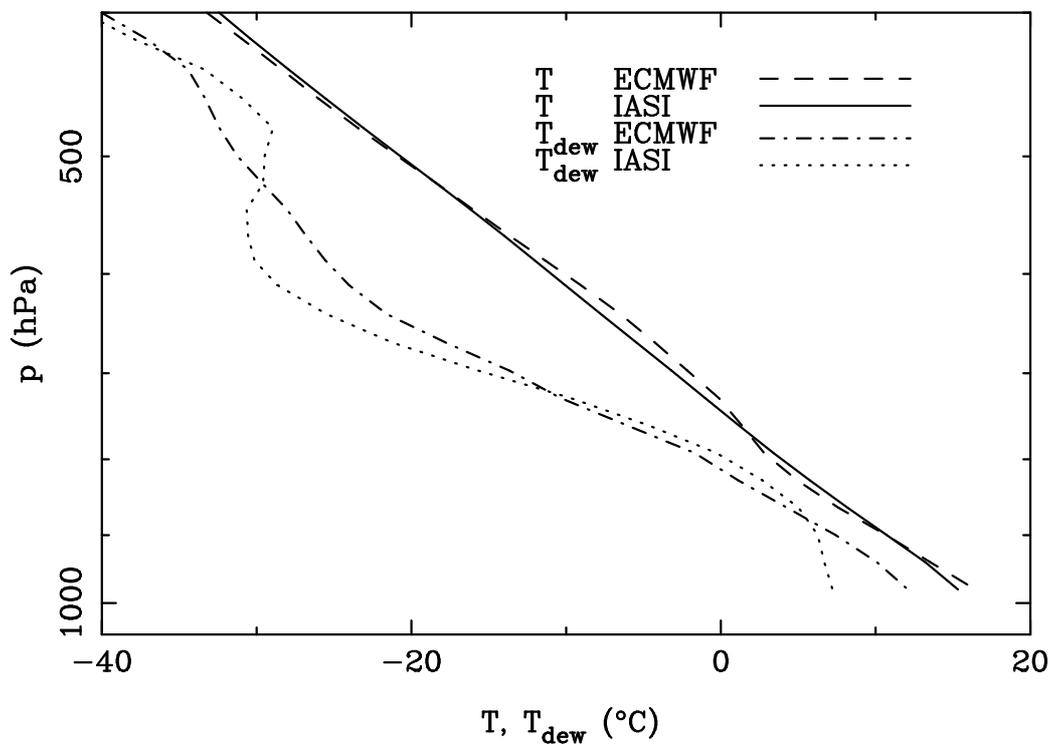}
\caption{\label{fig:oe2540}
Typical IASI retrieval is shown together with the
co--located ECMWF atmospheric profile. This one has 
a flatter temperature
profile, which is also relatively well retrieved, as well as the humidity
profile.}
\end{figure}

\begin{figure}
\noindent\includegraphics[angle=-90,width=\textwidth]{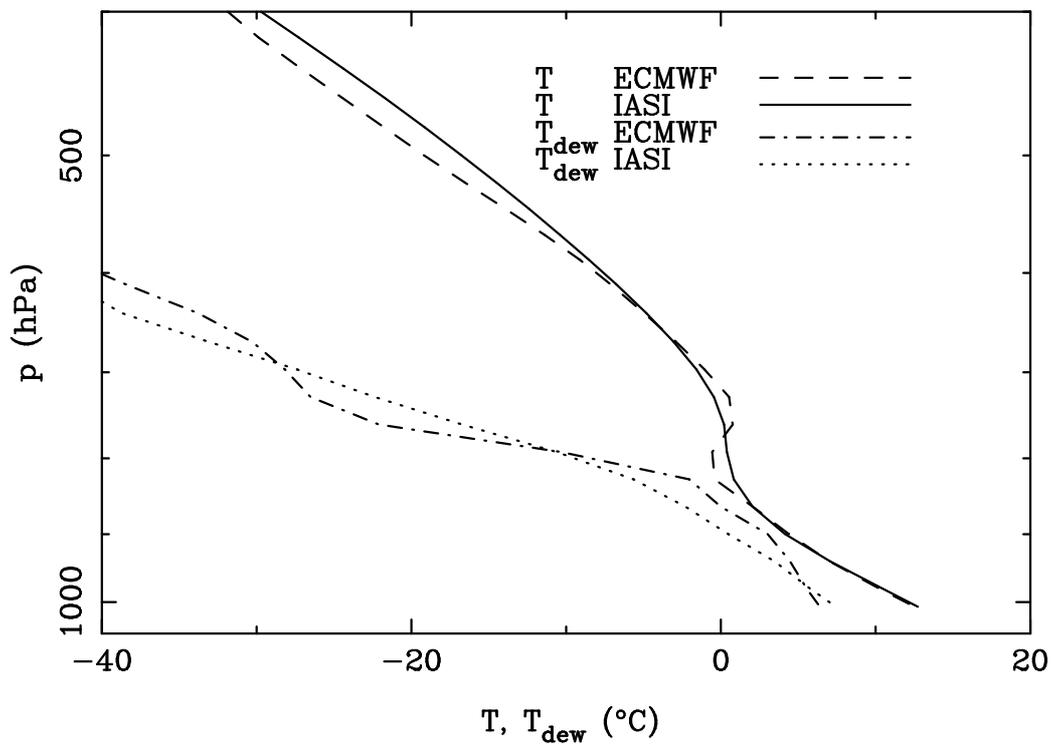}
\caption{\label{fig:oe3740}
Typical IASI retrieval is shown together with the
co--located ECMWF atmospheric profile.
Here we see how a strong low level inversion is also
reproduced by the retrieval even with high humidity at lower levels.}
\end{figure}

\begin{figure}
\noindent\includegraphics[angle=-90,width=\textwidth]{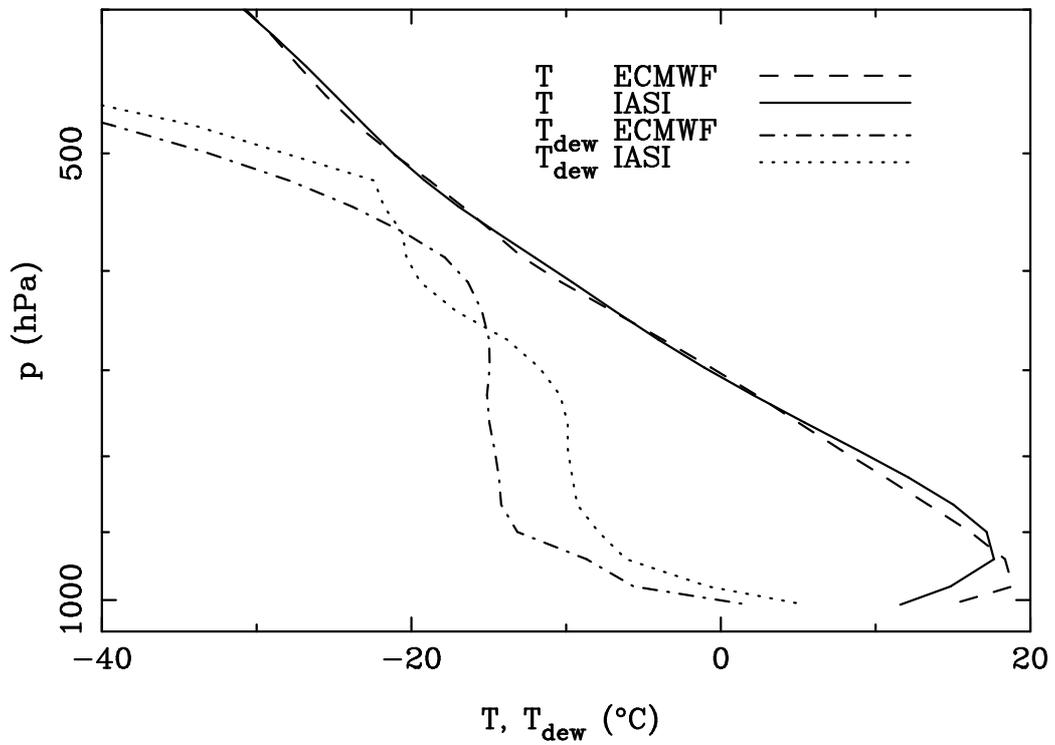}
\caption{\label{fig:oe4340}
Typical IASI retrieval is shown together with the
co--located ECMWF atmospheric profile.
In this figure a humidity maximum is well reproduced, this profile also has
a strong inversion near the surface.}
\end{figure}

A few examples of retrievals on non--polar, nighttime, clear sky,
 over the ocean scenes using the optimal error covariance matrix
are shown.
In Fig. \ref{fig:oe2221} a typical IASI retrieval is shown together with the
co--located ECMWF atmospheric profile. There is a low level inversion
that is clearly retrieved in this example. Also the humidity profile
is similar to the ECMWF analysis. In Fig. \ref{fig:oe2540} we have 
a flatter temperature
profile, which is also relatively well retrieved, as well as the humidity
profile. In Fig. \ref{fig:oe3740} we see how a strong mid level inversion is also
reproduced by the retrieval even with high humidity at lower levels.
In Fig. \ref{fig:oe4340} a humidity maximum is well reproduced, this profile also has
a strong inversion near the surface.

The global statistics of these 266 cases when using the optimal
error covariance matrix is shown in Fig. \ref{fig:stat_all} as a
solid line (OBS--CALC).
The IASI retrieval accuracy is between $0.9$ and $1.9\;{\rm K}$ in temperature
and below $6.5\;{\rm K}$ in humidity dew point temperature in the troposphere.
Note that these statistics have been computed directly
on RTIASI-4.1 pressure level grids without smoothing with the
averaging kernels. This implies a vertical
spacing between levels from $300$ to $400\; {\rm m}$  in the troposphere.

\subsection{Comparison with Radiosondes}
\label{sec:sondes}

\begin{figure}
\noindent\includegraphics[angle=-90,width=\textwidth]{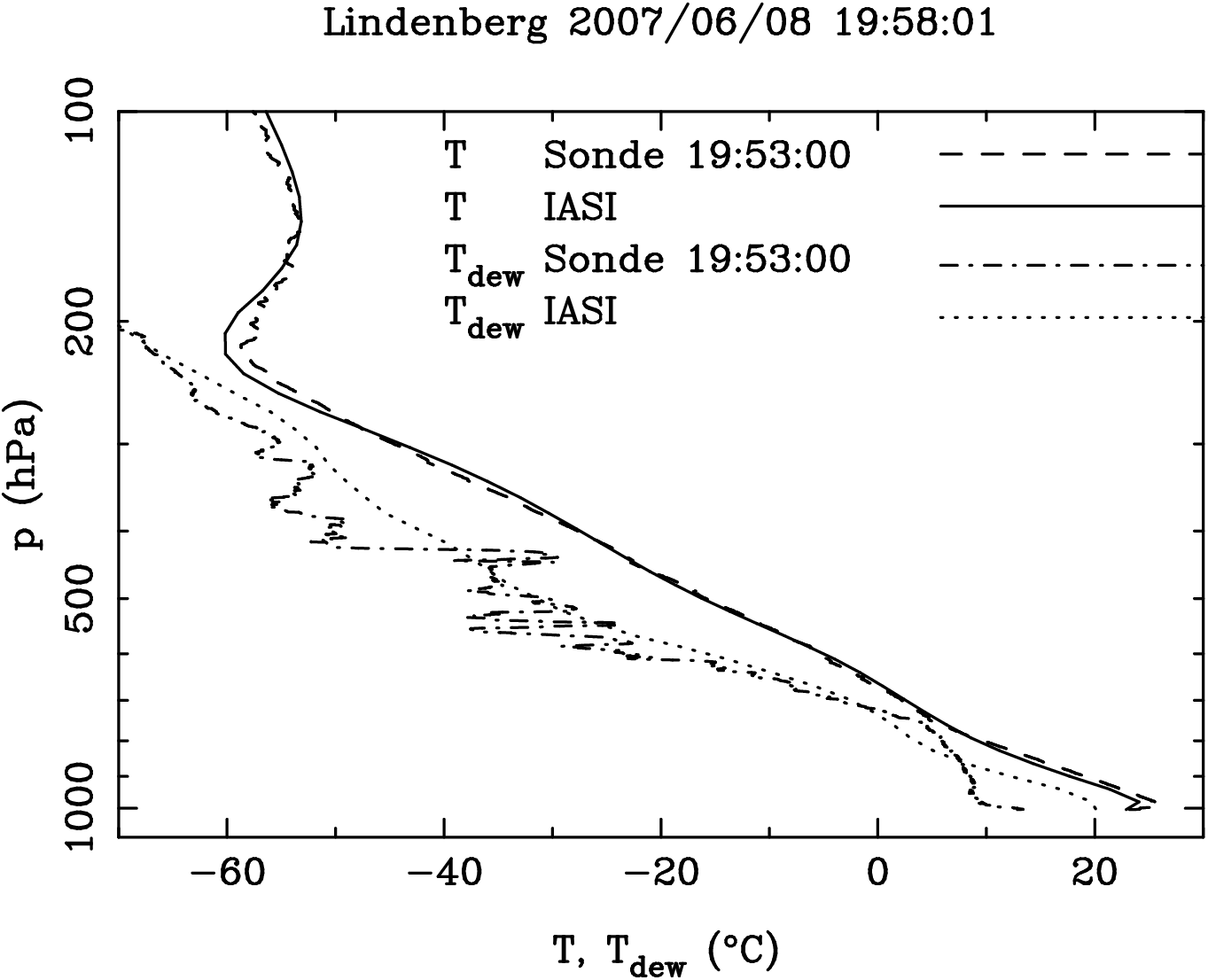}
\caption{\label{fig:lin0}
IASI retrieval fine tuned for ECMWF analyses compared with
co--located radiosondes from Lindenberg launched five minutes before
overpass time.}
\end{figure}

\begin{figure}
\noindent\includegraphics[angle=-90,width=\textwidth]{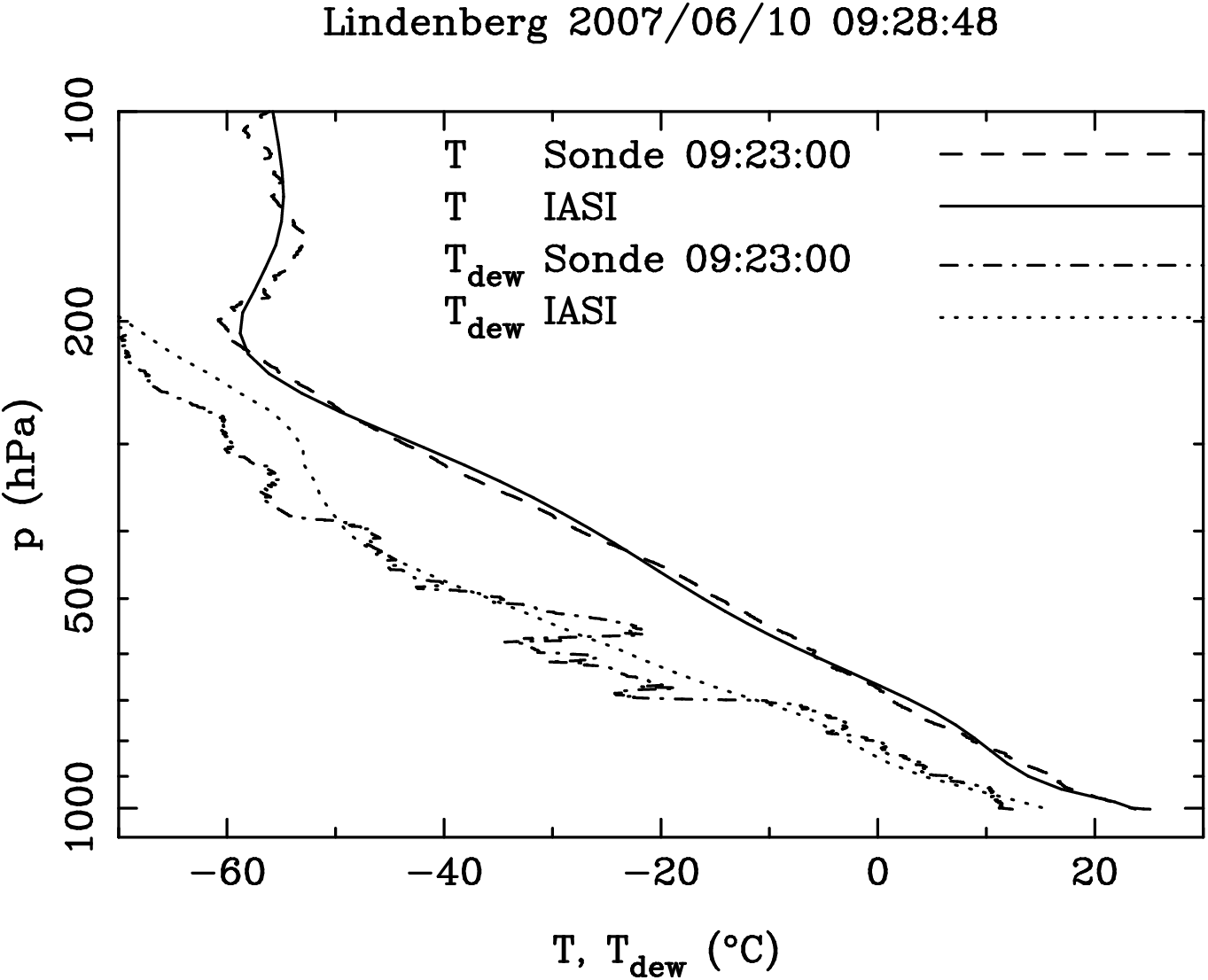}
\caption{\label{fig:lin1}
IASI retrieval fine tuned for ECMWF analyses compared with
co--located radiosondes from Lindenberg launched five minutes before
overpass time.}
\end{figure}

\begin{figure}
\noindent\includegraphics[angle=-90,width=\textwidth]{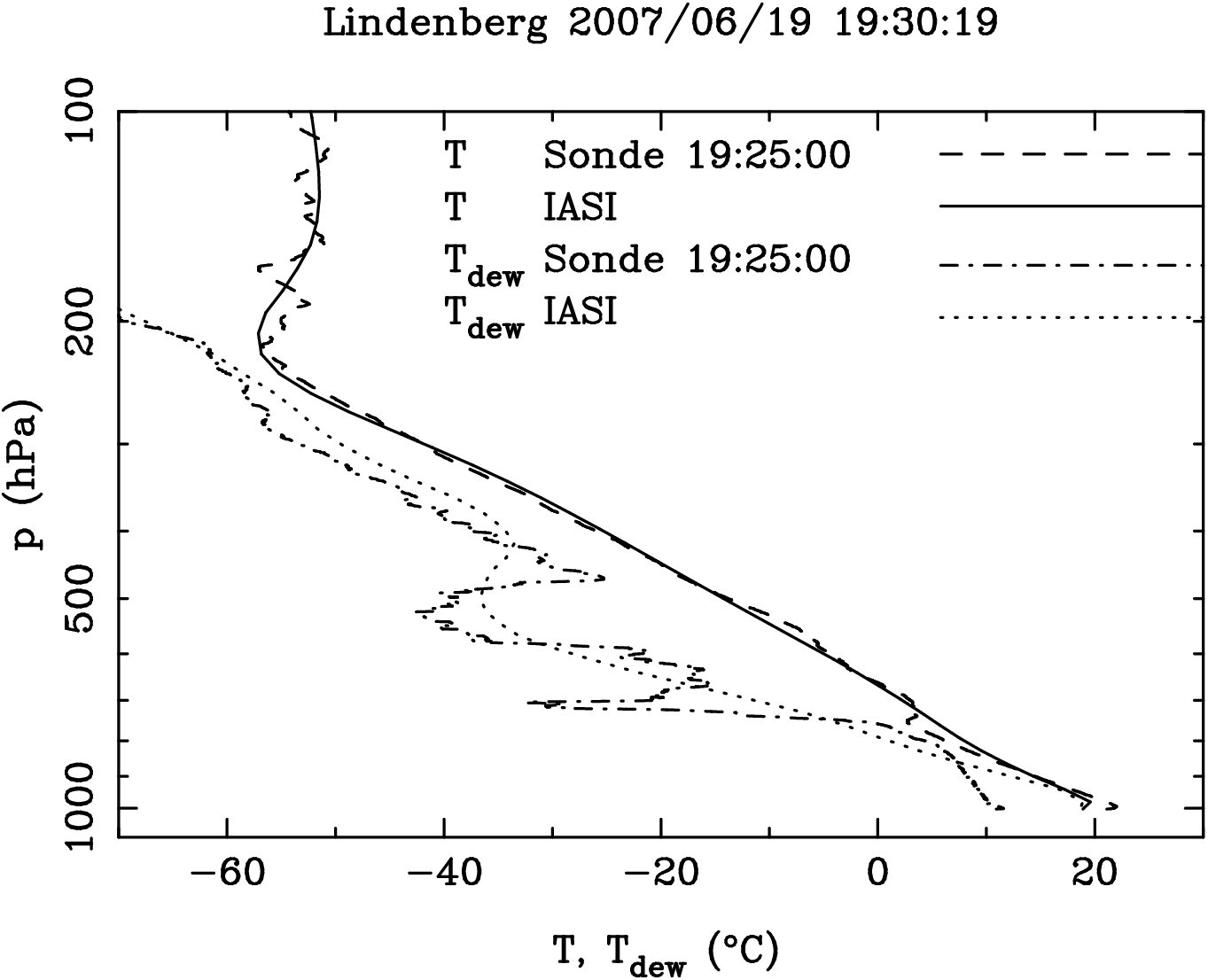}
\caption{\label{fig:lin2}
IASI retrieval fine tuned for ECMWF analyses compared with
co--located radiosondes from Lindenberg launched five minutes before
overpass time.}
\end{figure}

\begin{figure}
\noindent\includegraphics[angle=-90,width=\textwidth]{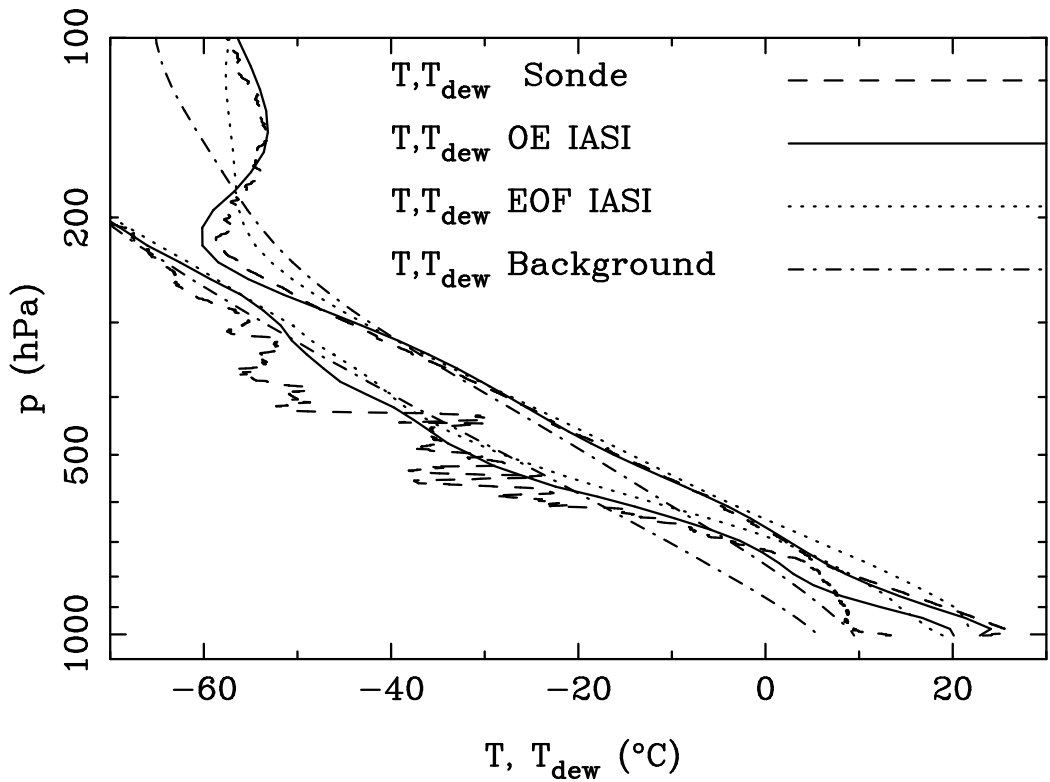}
\caption{\label{fig:lin0_eof_back}
IASI retrieval fine tuned for ECMWF analyses compared with
co--located radiosondes from Lindenberg launched five minutes before
overpass time. Also added in this figure are the first guess (EOF
retrieval) and background state of the optimal estimation retrieval.
Lines on the left and right side correspond to dew point temperatures and 
temperatures respectively.}
\end{figure}

Although the retrieval parameters have been fine tuned to ECMWF
analyses, they also compare well with co--located radiosondes.
In Fig. \ref{fig:lin0}, \ref{fig:lin1} and \ref{fig:lin2} we show 
three IASI retrievals together with their co--located
radiosondes launched five minutes
before overpass time from campaign data obtained
at Lindenberg. They were performed in clear sky situations at
night and daytime. It can be seen that the retrievals reproduce
particular interesting features of the atmosphere like
low level temperature inversions, levels of maximum humidity and
the tropopause. For illustrative purposes, the retrieval, sonde,
first guess (EOF retrieval) and background state for the first
of these examples is shown in Fig. \ref{fig:lin0_eof_back}.

\section{Discussion}
\label{sec:discussion}

It is normally the case that the radiative transfer modelling
of the atmosphere does not coincide exactly with the
observed infrared spectrum. Although these differences
may not seem to be very high, they are big enough
to degrade the retrievals significantly.
They can be caused by several reasons
like not knowing the exact concentration of trace gases,
erroneous line shapes in the radiative transfer model
or non--perfect atmospheric states, just to name a few.
There are usually two ways to correct for this error.
The first one of them is to model the atmosphere better
by either improving the radiative transfer model or by
using a more realistic atmospheric state vector, like
for example improving trace gases profiles.
The second is to bias correct
the observed radiances or brightness temperatures to match them
to the radiative transfer model ones.

Usually the errors when validating the retrievals are assumed
to come from three
different sources (Rodgers 1990):
instrument errors, radiative transfer
model errors and inacuracies in the representativeness
of the calibration dataset state vectors (NWP analyses in our case).
It is usually not simple to disentangle each one of these
sources of errors in the retrievals. In this paper
we have not tried to achieve this, but rather obtain
the best possible parameters (biases, error
covariance matrix and background properties) to
achieve the optimal retrievals when validated with the
validation dataset (NWP
analyses is the example shown here). In this way, we deal with all the errors
at once. This will make the method simple but at the
same time powerful by providing the best possible retrievals
when compared with one single source of validation dataset of 
atmospheric states.
The price we have to pay is that the retrievals are not the best
when compared with some ``ideal'' absolute reality of the atmosphere
because we will be overestimating the measurement error by including
an undesired source of error, the one from NWP analyses in this case.
In particular, by using more conservative values for the measurement
error covariance matrix the potential high vertical resolution
of IASI retrievals might be compromised.
It is clear that this method will be useful
when the errors of the calibration dataset of
atmospheric states compared to the absolute real ones 
are small enough for our purposes,
as could be the case here with temperature and
water vapor profiles coming from NWP analyses.
In any case, as we saw in Fig. \ref{fig:safe}, it is generally
safer to overestimate the measurement error, as we are doing
with this method, than to underestimate it.

Note that we are using a calibration and validation dataset
that does not represent the true atmospheric states perfectly
(ECMWF analyses), which in turn gives what
could be regarded as an oversized measurement error covariance
matrix as shown in Fig. \ref{fig:stdv}. Despite of this, the retrievals
do not seem to be extremely penalised in the vertical resolution
with respect to what could be expected. This can be
seen in Fig. \ref{fig:lin0}, where a very low level temperature inversion
is correctly reproduced.
The reason for this is that we are 
stopping the iterations of the optimal estimation
algorithm when the cost function does not descend significantly,
which in practice means that
the final brightness temperature residuals of the spectra are usually
well below
the values of the
measurement error covariance matrix, $S_\epsilon$.
More than the absolute values of $S_\epsilon$, 
the important parameters to be considered here are the relative
amounts within the $S_\epsilon$ matrix, which is what
effectively goes into the cost function in the optimal estimation.

One drawback of this technique is that we can only
use one source of atmospheric knowledge as the
calibration state vector. It would be advantageous
to extend this technique in such a way that more than one
source of measurements could be used, for example, using NWP analyses and
radiosondes at the same time.

A direct consequence of the analytical solution is that
there is one and only one measurement error covariance matrix
that is the optimal one for the validation dataset. If
this covariance is modified to better match some other
validation dataset or because we feel a lower value
would work better for the ``real'' atmospheric states,
we will have to settle with a degradation of the retrieval statistics
with respect to the first validation dataset.

The retrievals would be even closer to the real atmospheric
states if we used a better calibration dataset, like for
example radiosondes. Unfortunately, it is difficult to obtain
IASI co--located radiondes and there are not enough of them to make
a statistically signicant sample.

This method is the optimal one in the sense of providing the smallest
bias and standard deviation in the validation of the retrievals.
Other parameters different than these two could be devised to
characterize the goodness of the retrievals. In this case,
the analytical study could be modified to use these new parameters.

The method has been designed to make optimal retrievals.
It remains to be seen whether this same or other kind of similar analytical study
would also be useful for assimilation in NWP models. In this case, the
processing chain is much longer and does not stop in the retrievals
but extends much further up to the forecasts.

\appendix
\section{Analytical Proof of the Best Parameter Determination Method}
\label{sec:proof}

In this appendix we prove
that the optimal bias corrections and error covariance matrix to be used
in the retrievals are the OBS -CALC mean and covariance. We will show
this for the linearized forward radiative transfer model.

The retrieval method consists in minimizing a cost function, $J$, with
respect to $x'$. The cost function
can be explicitly written as,

\begin{equation}
 J = ( y' - F(x') )^T S_{\epsilon}^{-1} ( y' - F(x') ) + ( x' - x'_a )^T S_{a}^{-1} ( x' - x'_a ).
\end{equation}

Here we have used the usual matrix notation similar to that from Rodgers (2000),
being $x'$ the atmospheric state, $F$ the forward model, 
$y'$ the hyperspectral measurements,
$S_{\epsilon}$ the measurement error covariance matrix used in the retrieval,
$S_{a}$ the a--priori covariance matrix and
$x'_a$ the a--priori atmospheric state.

This complex non--linear problem is usually linearized by expanding the
forward model into a Fourier series around 
a reference point $x'_*$, which in general
will be different from the a--priori background state $x'_a$,

\begin{equation}
 y' \simeq y'_* + K ( x' - x'_* ),
\end{equation}

where $K$ is the Jacobian of the forward model $F$. We will define
$x$ and $y$ as the departures of the atmospheric states and measurements
from the reference point $x'_*$ and $y'_*$ respectively,

\begin{equation}
 y \equiv y' - y'_*,
\end{equation}

\begin{equation}
 x \equiv x' - x'_*.
\end{equation}

After the linearisation the cost function becomes,

\begin{equation}
 J = ( y - K x )^T S_{\epsilon}^{-1} ( y - K x) \; + \; (x - x_a)^T S_{a}^{-1} ( x - x_a ).
\end{equation}

In the retrieval process we try to minimize this function by making its derivative
equal to zero,

\begin{equation}
\frac{\partial J}{\partial x} \; = \; - K^T S_{\epsilon}^{-1} ( y - K x ) \; + \; S_a^{-1} ( x - x_a) \; + \; 
[\;]^T,
\end{equation}

where $[\;]^T$ denotes the transpose of the rest of the right hand side of the equation.
Solving for $x$ we obtain the familiar retrieval expression,

\begin{equation}
x_R = \left( K^T S_{\epsilon}^{-1} K + S_a^{-1} \right)^{-1} 
\left( K^T S_{\epsilon}^{-1} y_o +
S_a^{-1} x_a \right),
\end{equation}

where $x_R$ stands for the retrieved atmospheric state and we have substituted
$y_o$ for $y$ to stress that this is the observed spectrum. The reason
for this is to differentiate this spectrum from
the calculated one, $y_c$. The latter is obtained by applying the
radiative transfer model ($K$) to the calibration/validation
dataset of atmospheric states, $x_v$.

We now proceed to calculate the bias of the retrievals by comparing with
the validation dataset 
state of the atmosphere (NWP analyses for example), $x_v$,

\begin{eqnarray}
& x_R  - x_v = \nonumber \\
& \left( K^T S_{\epsilon}^{-1} K + S_a^{-1} \right)^{-1} 
\left( K^T S_{\epsilon}^{-1} y_o +
S_a^{-1} x_a \right) - x_v.
\end{eqnarray}

If we now multiply the $x_v$ term by $( K^T S_{\epsilon}^{-1} K + S_a^{-1} )^{-1}
( K^T S_{\epsilon}^{-1} K + S_a^{-1} )$, rearranging terms and taking
into account that $K x_v$ is what we have called the calculated spectrum,
$y_c$, we obtain,

\begin{eqnarray}
 x_R  - x_v = 
 \left[ K^T S_{\epsilon}^{-1} K + S_a^{-1} \right]^{-1} \cdot \nonumber \\
 \left[
K^T S_{\epsilon}^{-1} ( y_o - y_c ) +
S_a^{-1} ( x_a - x_v) \right].
\label{eq:xdiff}
\end{eqnarray}

By taking the expected value we obtain the final expression
for the bias,

\begin{eqnarray}
 \overline{x_R  - x_v} = 
 \left[ K^T S_{\epsilon}^{-1} K + S_a^{-1} \right]^{-1} \cdot \nonumber \\
 \left[
K^T S_{\epsilon}^{-1} ( \overline{y_o - y_c} ) +
S_a^{-1} ( \overline{x_a - x_v}) \right].
\label{eq:biasexpression}
\end{eqnarray}

We can see from this expression that the bias comes from two sources. The first
source is the OBS - CALC bias in the spectra ($\overline{y_o - y_c}$).
The second one comes from the difference between the background a--priori
 and the calibration atmospheric state. In order to minimize the bias in the
retrievals we should bias correct the observations with
the OBS - CALC average, such that in the end,

\begin{equation}
 \label{eq:bias}
\overline{y_c} = \overline{y_o}.
\end{equation}

Also the background
state should be equal to the average atmospheric states being
retrieved 

\begin{equation}
\label{eq:xa}
 x_a = \overline{x_v}.
\end{equation}

The error of the retrievals can be measured with the covariance between
the retrieved and the validation atmospheric profiles,

\begin{equation}
 {\rm Cov}(x_R-x_v) = ( x_R - x_v ) ( x_R - x_v)^T.
\end{equation}

If we now include the atmospheric profile difference
from Eq. \ref{eq:xdiff} we obtain,

\begin{eqnarray}
 {\rm Cov}(x_R-x_v) = \nonumber \\
\left[ K^T S_{\epsilon}^{-1} K + S_a^{-1} \right]^{-1} \cdot \nonumber \\
\left[ K^T S_{\epsilon}^{-1} ( y_o - y_c ) + S_a^{-1} ( x_a - x_v ) \right] \cdot \nonumber \\
\left[ ( y_o - y_c )^T S_{\epsilon}^{-1} K + ( x_a - x_v )^T S_a^{-1} \right] \cdot \nonumber \\
\left[ K^T S_{\epsilon}^{-1} K + S_a^{-1} \right]^{-1}.
\label{eq:covariance}
\end{eqnarray}

The optimal retrieval parameters, $S_{\epsilon}^{-1}$, $S_a^{-1}$ and
$x_a$ can be calculated by taking the derivative of this covariance with
respect to $S_{\epsilon}^{-1}$ and making it equal to zero,

\begin{eqnarray}
 \frac{\partial {\rm Cov}(x_R-x_v)}{\partial S_{\epsilon}^{-1}} =  \nonumber \\
\left[ K^T S_{\epsilon}^{-1} K + S_a^{-1} \right]^{-1} K^T \otimes K \left[ K^T S_{\epsilon}^{-1} K + S_a^{-1} \right]^{-1} \cdot \nonumber \\
\left[ K^T S_{\epsilon}^{-1} ( y_o - y_c ) + S_a^{-1} ( x_a - x_v ) \right] \cdot \nonumber \\
\left[ ( y_o - y_c )^T S_{\epsilon}^{-1} K + ( x_a - x_v )^T \right] \cdot \nonumber \\
\left[ K^T S_{\epsilon}^{-1} K + S_a^{-1} \right]^{-1} \nonumber \\
+ \nonumber \\
\left[ K^T S_{\epsilon}^{-1} K + S_a^{-1} \right]^{-1} \nonumber \\
K^T \otimes \left[ y_o - y_c \right] \nonumber \\
\left[ ( y_o - y_c )^T S_{\epsilon}^{-1} K + ( x_a - x_v )^T \right] \cdot \nonumber \\
\left[ K^T S_{\epsilon}^{-1} K + S_a^{-1} \right]^{-1} \nonumber \\
+ [ \; ]^T,
\end{eqnarray} 

where $[\;]^T$ denotes the transpose of the rest of the right hand side of the equation
and the symbol $\otimes$ represents the tensor product of two matrices in the sense
that each element of the four dimensional tensor product is,

\begin{equation}
 \left( A \otimes B \right)_{i,j,k,l} = A_{i,k} B_{l,j}.
\end{equation}

Note that the since the covariance matrices are symmetric,
$[S_{\epsilon}^{-1}]^T = S_{\epsilon}^{-1}$. The same applies for $S_a^{-1}$.

Rearranging terms and averaging over many cases we are left with,

\begin{eqnarray}
 \frac{\partial \overline{{\rm Cov}(x_R-x_v)}}{\partial S_{\epsilon}^{-1}} =  \nonumber \\
\left[ K^T S_{\epsilon}^{-1} K + S_a^{-1} \right]^{-1} K^T \otimes \lbrace \nonumber \\
-K \left[ K^T S_{\epsilon}^{-1} K + S_a^{-1} \right]^{-1} K^T S_{\epsilon}^{-1} \cdot \nonumber \\
\overline{(y_o-y_c) (y_o-y_c)^T} S_{\epsilon}^{-1} K \nonumber \\
+ \overline{(y_o - y_c) (y_o-y_c)^T} S_{\epsilon}^{-1} K \nonumber \\
-K \left[ K^T S_{\epsilon}^{-1} K + S_a^{-1} \right]^{-1} K^T S_{\epsilon}^{-1} \overline{(y_o-y_c) (x_a - x_v)^T} S_a^{-1} \nonumber \\
-K \left[ K^T S_{\epsilon}^{-1} K + S_a^{-1} \right]^{-1} S_a^{-1} \overline{(x_a - x_v) (y_o-y_c)^T} S_{\epsilon}^{-1} K \nonumber \\
-K \left[ K^T S_{\epsilon}^{-1} K + S_a^{-1} \right]^{-1} S_a^{-1} \overline{(x_a - x_v)  (x_a - x_v)^T} S_a^{-1} \nonumber \\
+\overline{(y_o - y_c) (x_a - x_v)^T} S_a^{-1} \nonumber \\
\rbrace \left[ K^T S_{\epsilon}^{-1} K + S_a^{-1} \right]^{-1} \nonumber \\
+ [ \; ]^T. \nonumber \\
\label{eq:covfinal}
\end{eqnarray} 

We will now show that this derivative is zero if we set,

\begin{eqnarray}
\label{eq:se}
S_{\epsilon} = \overline{(y_o - y_c) (y_o - y_c)^T} \\
\label{eq:sa}
S_a = \overline{(x_a - x_v)(x_a - x_v)^T}.
\end{eqnarray}

Introducing Eqs. \ref{eq:se} and \ref{eq:sa} into Eq. \ref{eq:covfinal} we can certify
that all terms including only $x$'s or $y$'s of Eq. \ref{eq:covfinal} vanish,
leaving only $x$ and $y$ cross--product terms.


Let us now analyze the cross--product term and show that it is also zero,

\begin{equation}
 \overline{(y_o - y_c) (x_a - x_v)^T} = \overline{(y_o - y_c)} x_a^T - \overline{y_o x_v^T} + 
\overline{y_c x_v^T} \;
\label{eq:crossproduct}
\end{equation}

The first term of the right hand side is zero when we apply the bias corrections
from Eq. \ref{eq:bias}.
If the forward model, $F$, reproduces well enough the properties of the real
atmosphere, the calculated spectra should have similar statistical properties
as the observed one in the sense that their cross--covariances with
the validation atmospheric states should be similar,
 
\begin{equation}
 \overline{y_o x_v^T} \approx \overline{y_c x_v^T},
\end{equation}

which implies that the last two terms of the right hand side of Eq. \ref{eq:crossproduct}
are also approximately zero.

This concludes the proof obtaining as results Eqs. \ref{eq:bias}, \ref{eq:xa}, \ref{eq:se} and \ref{eq:sa}.
Using these solutions we can also calculate the final error retrieval covariance matrix to obtain,

\begin{equation}
{\rm Cov}(x_R-x_v) = \left[ K^T S_{\epsilon}^{-1} K + S_a^{-1} \right]^{-1},
\end{equation}

which is the usual accepted expression (Rodgers 2000).

\clearpage

%
%
%



\clearpage

\begin{table}
\begin{flushleft}
\caption{Scene selection.\label{tab:scene}}
\begin{tabular}{c}
\hline
Cloud detection\\
\hline
$-1~{\rm K} < T(3.9~\mu {\rm m}) - T(10.8~\mu {\rm m})$
\footnotemark{a}
$< 3~{\rm K}$ \\
$T(10.8~\mu {\rm m}) > 276~{\rm K}$ \\
$T(11.0~\mu {\rm m}) > SST$
\footnotemark{b}
$- 2.2~{\rm K}$ \\
$T(4.0~\mu {\rm m})-T(11.0~\mu {\rm m}) > 12~{\rm K}$ \\
$T(9.3~\mu {\rm m})-T(11.0~\mu {\rm m}) < 0~{\rm K}$ \\
$T(11.0~\mu {\rm m})-T(12.0~\mu {\rm m}) < 1~{\rm K}$ \\
$T(11.0~\mu {\rm m})-T(13.6~\mu {\rm m}) > 18~{\rm K}$ \\
\hline
Others\\
\hline
$|Solar\;zenith\;angle| < 80^\circ$\\
$|Latitude| < 50 ^\circ$\\
$|Scan\;angle| < 15 ^\circ$\\
\hline
\end{tabular}
\end{flushleft}
\footnotetext{a}{$T(10.8~\mu {\rm m})$, for example, is the brightness
temperature of an AIRS channel that lies
in that wavelength ($10.8~\mu {\rm m}$).}
\footnotetext{b}{SST is the sea surface temperature
derived from ECMWF analysis.}
\end{table}

\end{document}